\documentclass{aa}
\usepackage{txfonts}
\usepackage{graphicx}
\usepackage{psfrag}
\usepackage{natbib}
\usepackage{lscape}
\usepackage{multirow}

\begin{document}
\title{Deep, wide-field, global VLBI observations of the Hubble Deep
 Field North and Flanking Fields}
\subtitle{}
\author{S. Chi (deceased) \inst{1, 2, 3}
            \and P. D. Barthel\inst{1}
            \and M. A. Garrett \inst{3, 4, 5} }
\titlerunning{HDF-N and HFF}
\authorrunning{S. Chi et al.}
\offprints{P. D. Barthel\\  \email{pdb@astro.rug.nl}}
\date{Received November 23, 2012; accepted December 21, 2012}
\institute{Kapteyn Astronomical Institute, University of Groningen,
  P.O.~Box 800, 9700 AV Groningen, The Netherlands  \and
  Joint Institute for VLBI in Europe (JIVE), P.O.~Box 2, 7990 AA Dwingeloo,
  The Netherlands  \and
  Netherlands Foundation for Research in Astronomy (ASTRON), P.O.~Box 2,
  7990 AA Dwingeloo, The Netherlands \and
  Centre for Astrophysics and Supercomputing, Swinburne University of Technology,
   Mail number H39, P.O.~Box 218, Hawthorn, Victoria 3122, Australia
   \and Leiden Observatory, P.O.~Box 9513, 2300 RA Leiden, The Netherlands}

\abstract
 {Dust is commonly present in weakly radio emitting star-forming galaxies
  and this dust may obscure the signatures of accreting black holes in
  these objects.}
 {We aim to uncover weak active galactic nuclei, AGN, in the faint radio source
  population by means of deep high-resolution radio observations.}
 {VLBI observations with a world-wide array at unparallelled sensitivity
  are carried out to assess the nature of the faint radio source population
  in the Hubble Deep Field North and its flanking fields.}
 {Images of twelve compact, AGN-driven radio sources are presented. These represent
  roughly one quarter of the detectable faint radio source sample. Most, but not
  all of these low power AGN have X-ray detections.}
 {The majority of the faint radio source population must be star-forming
  galaxies. Faint AGN occur in a variety of (distant) host galaxies, and these
  are often accompanied by a dust-obscured starburst. Deep, high-resolution VLBI
  is a unique, powerful technique to assess the occurrence of faint AGN.}

\keywords{galaxies: active -- galaxies: radio continuum --  galaxies: starburst}

\maketitle

\section{Introduction}

Deep wide-field radio observations of the Hubble Deep Field North
(HDF-N) and surrounding Flanking Fields (HFFs) have revealed a
population of sub-mJy and microJy radio sources, of which a substantial
fraction are associated with relatively distant star-forming galaxies
\citep{R98, garrett01, muxlow05}.  Optical studies of these fields
indicate that a significant fraction of the hosts of these radio sources
are faint, remaining undetected in deep {\it Hubble Space Telescope}
$I$-band imaging \citep{richards99}.  Some of these objects can be seen
in the near-IR and have very red colours, leading to their
interpretation as distant, dust-enshrouded starbursts
\citep{dickinson00}. 

While X-ray and follow-up surveys have established a substantial surface
density of AGN among these and other distant galaxies
\citep[e.g.][]{alexander03, bauer04, xue11}, the actual AGN surface (and
space) density is most likely higher.  Many models for the growth of
supermassive black holes predict that the most rapid growth phase will
be obscured by Compton-thick absorption \citep{fabian99, hopkins06},
which is for instance supported by mid-infrared observations of Type-2
QSOs \citep{martinez05}.  The importance of strongly obscured AGN became
apparent from studies of radio-excess (with respect to infrared) AGN
without X-ray counterparts \citep{donley05}, and from the X-ray
properties of seemingly normal or optically obscured star-forming
galaxies displaying excess infrared emission \citep{daddi07, fiore09,
murphy09, pope08, georga10, luo11, alex11}.  Infrared and submm
photometric surveys have proven extremely valuable.  The former permit
AGN identification on the basis of the shape of the infrared SED
\citep[e.g.][]{alonso06, donley07}, or using infrared color-color
selection \citep[e.g.][]{stern05, lacy04, donley12}.  The latter survey
technique finds distant starburst galaxies with an occasional embedded
AGN \citep[e.g.][]{pope06, lutz10, biggs10}.  The general reliability of
these selection techniques, and the level of AGN completeness
nevertheless remain debatable. 

Access to the complete infrared-submm spectral energy distributions,
SEDs, has recently become possible through {\it Herschel Space
Observatory} data, and the SED data permit separation of AGN- and
starburst-heating in a consistent fashion.  On the basis of extensive
{\it Herschel} observations, redshift evolution of star-formation in a
large sample of X-ray selected AGN hosts was explained by
\citet{mullaney12} as being due to evolution of the specific
star-formation rate, sSFR, in otherwise normal star-forming host
galaxies.  Employing multi-wavelength SED decomposition, a frequent
occurrence of excess radio emission attributed to weak AGN in distant
star-forming galaxies was very recently established by \citet{moro12}. 

One of the key questions in extragalactic astronomy is the {\it true}
incidence of AGN in distant starburst galaxies and the symbiotic
interaction of the two phenomena \citep[e.g.][]{silk05, schweitzer06}. 
All abovementioned studies make it very clear that the phenomena of
star-formation and accretion-driven nuclear activity can occur in
concert.  It is however also very clear that dust obscuration plays an
important role, preventing full and complete assessment of the
occurrence of AGN among mildly or strongly star-forming galaxies using
ultraviolet, optical, infrared and even X-ray observations. 

Dust obscuration does not affect the radio band, hence cm-wavelength
radio emission at arcsec scale resolution can be used to find (weak) AGN
among star-forming galaxies\footnote{Radio-loud AGN obviously stand out
in radio surveys through their radio morphology and luminosity.}.  The
radio-infrared ratio can be used to isolate radio-excess objects among
the faint radio source population, just on itself
\citep[e.g.][]{donley05}, or in combination with radio morphology and
spectral shape \citep{seymour08}.  Considering the nature of the faint
radio source population, the power of subarcsec radio imaging was shown
by \citet{muxlow05}, and in the case of distant infrared galaxies by
\citet{casey09}.  The present study follows up on the work of
\citet{muxlow05}, employing Very Long Baseline Interferometry, VLBI, to
obtain 1--10 milliarcsec scale resolution on the 92 faint radio sources
reported by these authors in the HDF-N and its flanking fields.  Only
such deep VLBI observations are capable of distinguishing between radio
emission generated by star-formation processes and AGN activity in
distant dust-obscured systems. 

Previous VLBI observations of the HDF-N, carried out with the European
VLBI Network, EVN \citep{garrett01} already demonstrated the power of
deep, high resolution VLBI imaging in discriminating between starburst
and AGN activity in dust obscured systems, yielding important
morphological information as well as brightness temperatures.  This
paper presents the culmination of that study, employing a global array
of VLBI telescopes attaining milliarcsec-scale resolution and
micro-Jansky sensitivity over a considerably larger field. 

\section{Global VLBI observations and data analysis}

Observations of the HDF-N and HFF region were made with a global VLBI
array on 20--22 February 2004 at 1.4\,GHz.  A total of 36hrs observing
time was split into three 12hr runs.  The global VLBI array consisted of
16 telescopes in Europe and the USA, including the 100-m Effelsberg,
100-m Green Bank, and 76-m Lovell (Jodrell Bank) telescopes, recording
data at a bit rate of 128 Mbits/s (2$\times$8\,MHz bands) in both right
(R)- and left (L)-hand circular polarisation.  The observations were
made in phase-referencing mode with a typical cycle time of 160sec on
the secondary phase-calibrator and 330sec on the HDF-N.  The total
on-source integration time on the HDF-N was $\sim$19 hours.  Two
phase-calibrators were used: a strong primary calibrator J1241+602 (a
compact $S_{1.4 ~\mathrm {GHz}}\sim 455$ mJy source lying 2$\degr$ from
the HDF-N) and a fainter secondary calibrator J1234+619 ($S_{1.4
~\mathrm{GHz}}\sim 20$ mJy lying only 20$\arcmin$ from the HDF-N).  The
primary calibrator was typically observed for 100sec once every
$\sim$42min; more frequent scans on the nearby, secondary calibrator
permitted to make the finer phase corrections.  \\

The data were processed at the European VLBI Network (EVN) correlator at
the Joint Institute for VLBI in Europe, JIVE, in Dwingeloo, the
Netherlands.  Each 8\,MHz band (R\&L) was correlated in separate passes
in order to maximise spectral resolution (256 channels/IF) thereby
minimising bandwidth smearing.  The total data set was 675 Gbytes in
size making this one of the largest and most complex VLBI data sets ever
processed.  An integration time of 0.25seconds was employed to reduce
the effects of time smearing.  The phase center was coincident with a
470$\mu$Jy radio source, VLA J123642+621331 \citep{R00}, located just
outside the HDF-N (in an adjacent HFF; see Sect.3.1.1).  The data from
each IF were averaged, edited, and calibrated with the NRAO \verb"AIPS"
package, and gain parameters were subsequently applied to the unaveraged
original data.  The visibility amplitudes were calibrated using the
system temperatures and gain information provided by each telescope.  \\

In order to image out the entire HDF-N and HFF, the wide-field imaging
technique was used \citep{garrett99}.  A large number (92 -- see below)
of dirty images and dirty beams for each 8 MHz band (R\&L
simultaneously) were generated using IMAGR, and co-added; fields with a
clear detection were subsequently CLEANed using APCLN.  The absolute
flux density scales of the VLBI observations are expected to be better
than 5\%.  However, we note that the VLBI flux densities have not been
corrected for primary beam attenuation.  The calibrators used have
positions that are measured to better than one milliarcsec.  We
therefore estimate the uncertainties in VLBI positions to be set by
errors introduced by the ionosphere.  With a separation of 0.5 degrees
between our main phase reference source and the target fields, we
estimate an error of $\sim$1 -- 2 milliarcsec in each coordinate.  \\

\section{Global VLBI detections in the HDF-N and HFF region}

The wide-field global VLBI survey covered a total of 201 arcmin$^{2}$
divided into four annular fields with different sensitivities and
angular resolutions.  The observations are both deeper and wider than
any previous VLBI observations of the field, attaining an r.m.s.  noise
level of 7.3 $\mu$Jy/beam with 4 mas (milli-arcsec) angular resolution
in the central 0$-$2$'$ part.  The outer annuli, at 2$'-$4$'$,
4$'-$6$'$, and 6$'-$8$'$ from the phase center, have 14$-$37
$\mu$Jy/beam r.m.s.  noise level and 19$-$27 mas angular resolution. 
Radio images were made for all of the 92 radio sources which were
detected by the MERLIN-VLA survey \citep{muxlow05} in these areas. 
These 92 sources had been classified on the basis of their arcsec-scale
radio morphologies combined with their infrared flux densities, as
follows: 18 AGN or AGN candidates, 3 starbursts with embedded AGN, 45
starbursts or starburst candidates, and 26 unclassified radio sources
\citep{muxlow05}.  Taking the variable VLBI sensitivity over the whole
field into account, 48 out of these 92 radio sources are bright enough
to be detected by our observations.  However, VLBI detection critically
depends on the angular dimension of the radio component(s): extended
arcsec-scale emission will be completely resolved by the small VLBI
beam. 

Our new global VLBI observations clearly reveal 12 compact radio sources
in the HDF-N and HFF above the local 5$\sigma$ detection level.  This
corresponds to a $\sim$25$\%$ detection rate.  Figure~1 presents the
radio-optical overlay of the fields, combining the arcsec and
milliarcsec radio imaging.  The global VLBI detections are listed in
Table~1, with the MERLIN-VLA radio structure description and object
classification in columns 10 and 11.  The radio structural description
is as follows: C: compact, at 0.2~arcsec resolution; CE: compact with
2-sided extended emission; C1E: compact with 1-sided extended emission;
FR1: Fanaroff and Riley Class~1 edge-darkened double-lobed radio galaxy. 
The table also contains $q_{\rm 24}$-values, defined as the logarithm of
the 24$\mu$m/1.4\,GHz flux density ratio (where the integrated arcsec
scale radio emission as measured with the VLA was used in the
computation).  Among the VLBI sources, three (VLA J123642+621331, VLA
J123644+621133, and VLA J123646+621404) had been detected in earlier EVN
1.6\,GHz observations \citep{garrett01}, but the other nine radio
sources are new.  All detected objects are described separately, in two
Subsections below. 

\begin{figure*}[!h]
   \centering
   \resizebox{\textwidth}
             {!}{\includegraphics[clip=true]{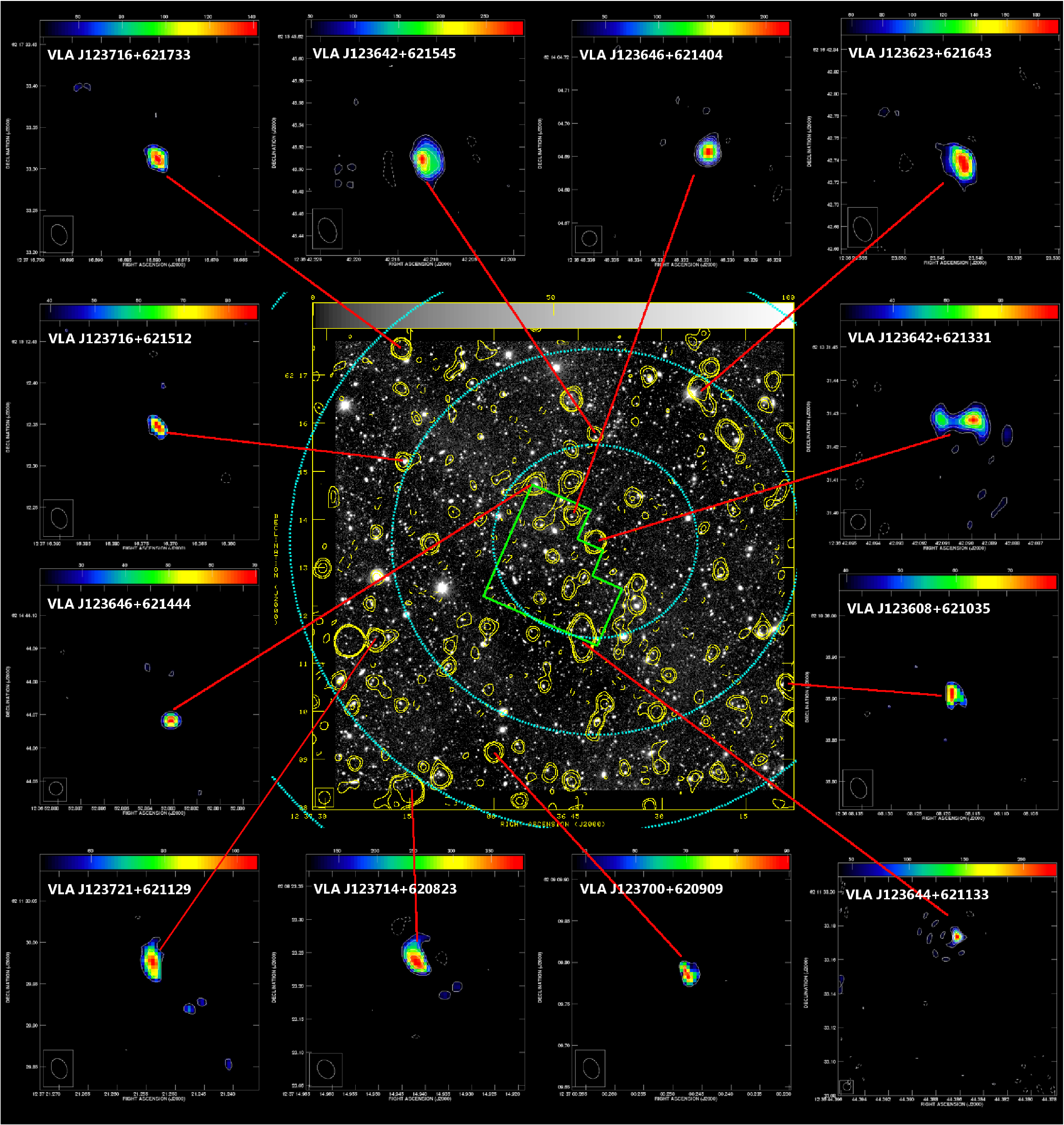}}
      \caption{Composite image of the radio (WSRT 1.4\,GHz)--optical overlay 
               image of the HDF-N and HFF, surrounded by postage stamp images
               of the twelve compact VLBI-detected radio sources.The 
               cyan circles represent annuli of decreasing resolution and 
               sensitivity, and are drawn at 2, 4, 6, and 8 arcmin radius
               w.r.t. the phase center which coincides with radio AGN
               VLA J123642+621331 (see text Section~3).}
    \label{Fig1}
\end{figure*}

\subsection{Inner HDF-N and adjacent HFF}

\subsubsection{VLA J123642+621331}

VLA J123642+621331 lies just outside the HDF-N, in an adjacent HFF.  It
has a steep radio spectrum ($\alpha=0.94, ~S_{\nu}\propto
\nu^{-\alpha}$); its optical counterpart has $I=25$ \citep{R98,
muxlow05}.  HST NICMOS 1.1 $\mu$m (F110W), 1.6 $\mu$m (F160W), and KPNO
4-m $K$-band imaging detect a red host ($I_{814}-K = 2.0\pm 0.2$) for
the radio source, and spectra obtained by the Keck II telescope show a
single strong Ly$\alpha$ emission line at $z = 4.424$
\citep{waddington99, dickinson98}.  This source is also detected in the
{\it ISO} 15 $\mu$m and {\it Chandra} soft X-ray bands \citep{aussel99,
brandt01}.  \citet{waddington99} interpret the object as a dust-obscured
starburst with an embedded AGN. 

The deep MERLIN-VLA observations at 1.4\,GHz \citep{muxlow05} resolved
VLA J123642+621331, showing extended emission adjacent to an unresolved
core, which corresponds to the core detected by the EVN 1.6\,GHz
observations \citep{garrett01}.  The ratio of its radio and FIR
luminosity indicates that the object has a modest radio excess
\citep{garrett02}, providing further evidence for an embedded AGN: its
$q_{\rm 24}$-value is $-0.35$. 

The deep, high-resolution global VLBI image removes all doubts,
revealing a jet-like extension ($\sim$ 120 $\mu$Jy) emanating from a
compact AGN core.  The integrated flux density is in agreement with the
\citet{garrett01} measurement.  The separation between the jet and the
AGN core is about 70~pc.  This may be an example of a high redshift
ULIRG in which the high star-formation rate and the efficiency are
enhanced by AGN jet activity \citep[e.g.][]{silk05}. 

\begin{figure}[h]
   \centering
   \includegraphics[width=8.5cm]{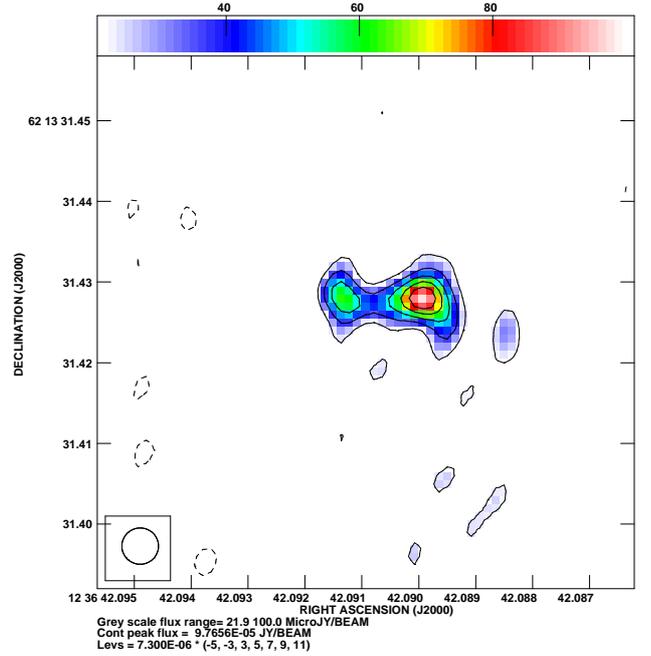}
      \caption{VLA J123642+621331. In this and subsequent radio images, the
        clean beam FWHM is specified in the
        bottom left corner. Contours are plotted starting at 3$\sigma$ and
        the relevant noise figure can be found below each image. The colour
        coding is specified above each image. }
      \label{Fig2}
\end{figure}

\subsubsection{VLA J123644+621133}

VLA J123644+621133 is optically identified with an $I=20.9$, very red
elliptical galaxy at $z=1.050$ \citep{cohen00}.  Its large-scale radio
structure and luminosity indicate a classical FR-I radio galaxy
\citep{R98, muxlow05}.  The AGN nature also follows from its negative
$q_{\rm 24}$ upper limit.  The previous EVN 1.6\,GHz observations
\citep{garrett01} revealed a compact core component and an additional
5$\sigma$ component located $\sim$60mas south of the core component, of
which the direction coincides with the extension of the VLA-only
\citep{R98} and MERLIN-VLA radio morphology \citep{muxlow05}.  The WSRT
and more recent \citep{M10} VLA flux densities indicate substantial
resolution effects.  Our high resolution VLBI observations merely detect
the core. 

\begin{figure}[h]
   \centering
   \includegraphics[width=8.5cm]{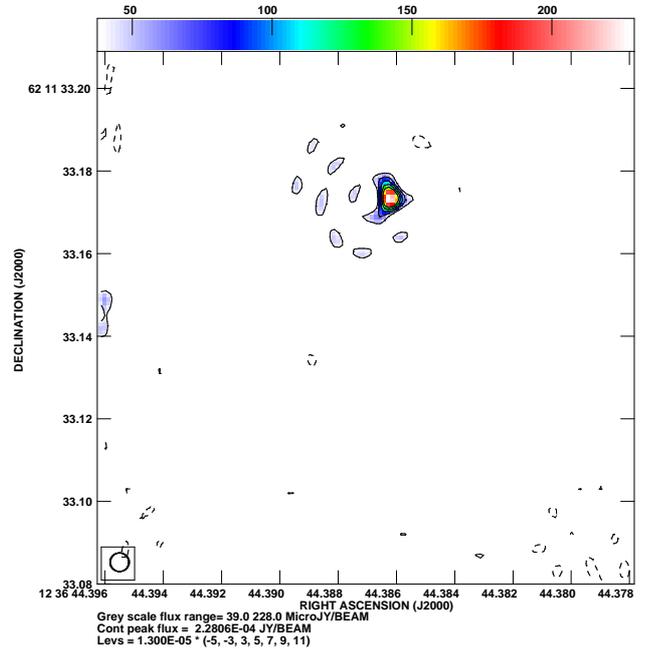}
      \caption{VLA J123644+621133}
      \label{Fig3}
\end{figure}

\subsubsection{VLA J123646+621404}

VLA J123646+621404 has a slightly inverted radio spectrum ($\alpha =
-0.04$) and is associated with a nearly face-on spiral galaxy
($I=20.7$), at a redshift of 0.961 \citep{cowie04}.  The MERLIN-VLA
image shows a compact component and two-sided extended emission, and EVN
1.6\,GHz observations detect a 4$\sigma$ radio source at this position
\citep{garrett01,muxlow05}.  Its radio variability is evident from
\citet{R98}, from the present data, and from recent VLA data
\citep{M10}.  Those VLA data in fact imply $q_{24} = -0.18$.  Taken
together with the broad emission line character \citep{phillips97,
brandt01}, there is no doubt about the presence of an AGN. 

\begin{figure}[h]
   \centering
   \includegraphics[width=8.5cm]{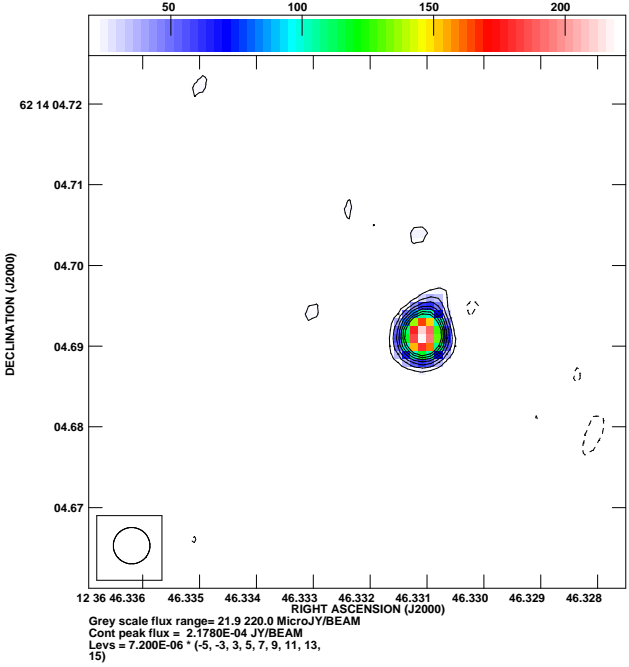}
   \caption{VLA J123646+621404}
   \label{Fig4}
\end{figure}

\subsubsection{VLA J123652+621444}

VLA J123652+621444 is an inverted spectrum radio source ($\alpha=-0.12$)
located in an adjacent HFF and observed to vary in intensity on the
timescale of months, suggesting the presence of an AGN
\citep{R98,R00,M10}.  The object displays radio excess, as judged from
its $q_{24}$-value.  MERLIN-VLA observations reveal a compact core,
which overlies the nucleus of an $I=18.5$ elliptical galaxy at
$z=0.321$, and a one-sided jet-like feature to the east
\citep{cowie04,muxlow05}.  The high resolution VLBI observations merely
detect the 0.08\,mJy radio core. 

\begin{figure}[h]
   \centering
   \includegraphics[width=8.5cm]{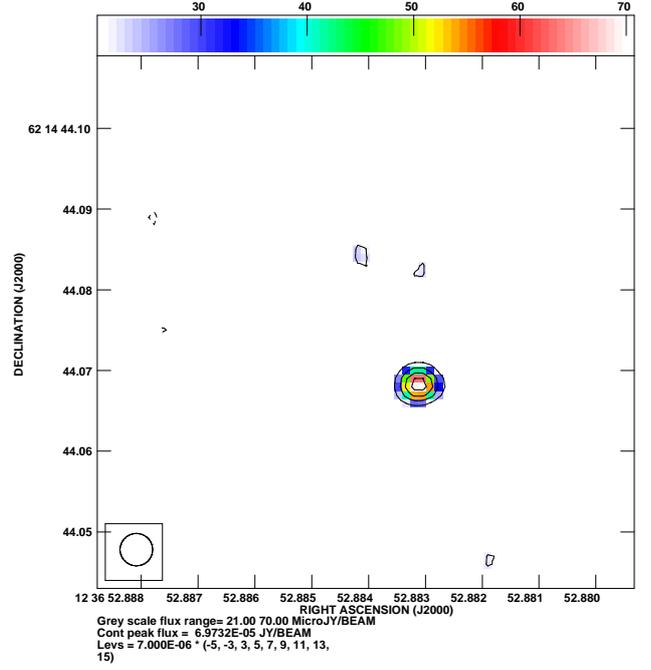}
   \caption{VLA J123652+621444}
   \label{Fig5}
\end{figure}

\subsection{Outer HFF region}

\subsubsection{VLA J123608+621035}

VLA J123608+621035 is a relatively flat-spectrum radio source
($\alpha=0.36$) hosted by an $I=20.6$ galaxy at $z=0.681$
\citep{R00,cowie04}.  The object is bright in the infrared, and falls on
the radio-FIR correlation.  Its compact radio core is straddled by
two-sided radio emission extending over $\la 1\arcsec$ \citep{muxlow05};
our VLBI detects a slightly extended 0.140mJy radio core. 
   
\begin{figure}[h]
   \centering
   \includegraphics[width=8.5cm]{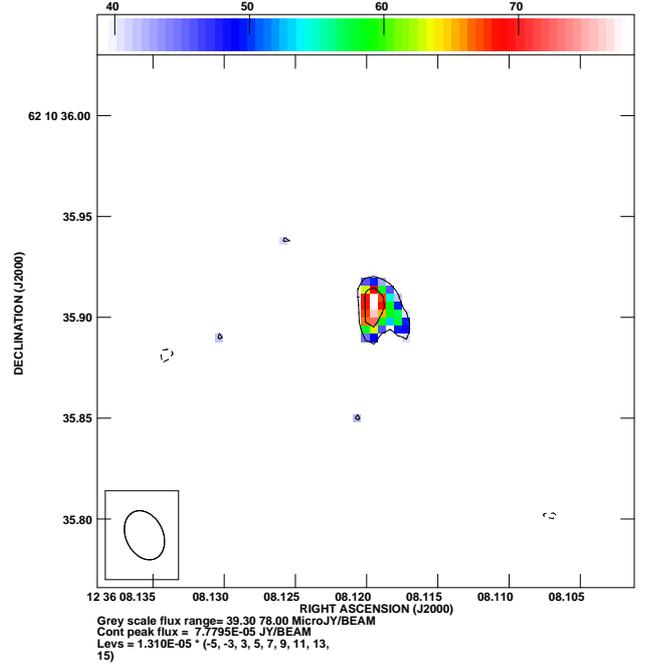}
   \caption{VLA J123608+621035}
   \label{Fig6}
\end{figure}

\subsubsection{VLA J123623+621642}

VLA J123623+621642 has a steep radio spectrum ($\alpha=0.63$) and its
MERLIN-VLA image shows a compact component with weak extended
(sub-arcsec) emission \citep{R00,muxlow05}.  The radio source is
identified with an $I=23.9$ galaxy at $z=1.918$; its radio and infrared
data indicate a radio-excess.  As inferred from the VLBI observations, a
substantial fraction of the radio luminosity is contained within its
milliarcsec-scale radio core. 

\begin{figure}[h]
   \centering
   \includegraphics[width=8.5cm]{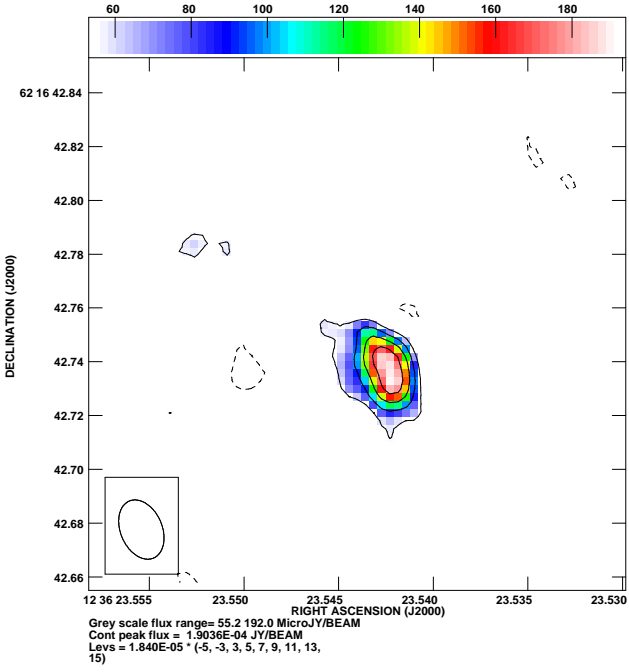}
   \caption{VLA J123623+621642}
   \label{Fig7}
\end{figure}

\subsubsection{VLA J123642+621545}

VLA J123642+621545 is an intermediate radio spectrum source
($\alpha=0.50$), associated with an $I=20.7$ galaxy at a redshift of
0.857 \citep{R00,hornschemeier01}.  The source was also detected by {\it
ISO} and {\it Chandra}.  \citet{muxlow05} suggest that the compact
component could be an AGN, but \citet{hornschemeier01} interpreted the
X-ray's as originating from a starburst system.  The VLBI observations
uncover a compact component of variable flux density in this
infrared-luminous object.  The object falls on the radio-infrared
correlation using the low-state VLA flux density, but definitely also
contains an AGN. 

\begin{figure}[h]
   \centering
   \includegraphics[width=8.5cm]{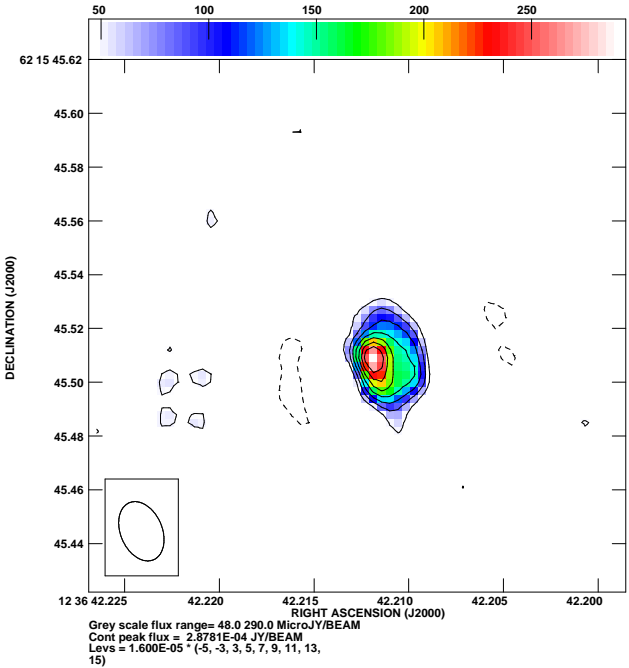}
   \caption{VLA J123642+621545}
   \label{Fig8}
\end{figure}

\subsubsection{VLA J123700+620909}

VLA J123700+620909 represents similar properties to VLA J123642+621331. 
It has relatively strong radio emission and a steep radio spectrum
($\alpha=0.89$).  There is no optical counterpart to $R = 27$
\citep{R98}.  The MERLIN-VLA observations detected the source, finding a
compact radio component and one-sided emission to the north-east.  At
the same time, deep {\it Spitzer} IRAC and MIPS imaging as well as SCUBA
850 $\mu$m observations \citep{pope06} also detected this source,
GN\,16.  Given its ERO-character as well as the results of the
multi-wavelength studies in sub-mm, IR, and radio, this source is
considered to be a starburst galaxy (SFR $\sim 1000 ~\mathrm
{M_{\odot}/yr}$), at $z_{\rm photom.} = 1.68$ \citep{pope05,pope06}. 
The VLBI observations find a $\sim 0.15$\,mJy compact radio core in this
mild radio-excess, $q_{24} = -0.08$ object. 

\begin{figure}[h]
   \centering
   \includegraphics[width=8.5cm]{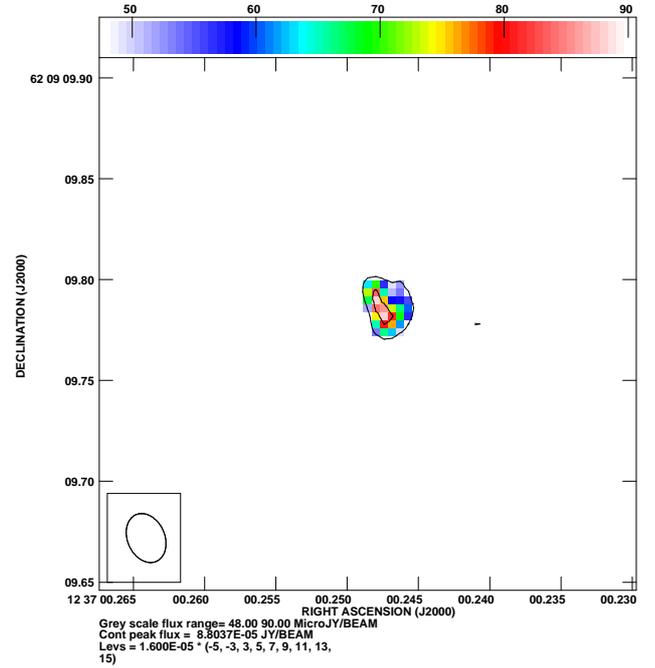}
   \caption{VLA J123700+620909, GN16}
   \label{Fig9}
\end{figure}

\subsubsection{VLA J123714+620823}

VLA J123714+620823 is associated with a faint, $R=24.25$ \citep{capak04}
red object.  Cowie (2010, priv.  communication) provided a tentative
redshift $z=0.847$.  With a total 1.4\,GHz flux density of well over a
mJy, the object has a moderate radio excess, as noted also by
\citet{donley05}.  The VLBI radio source is somewhat extended, and the
WSRT and VLA measurements indicate extended emission on the arcsec and
subarcsec scale.  The 21$\mu$Jy 8.4\,GHz VLA source J123714+620822
\citep{R98} is 0.3~arcsec away and its relation to J123714+620823 is
unclear. 

\begin{figure}[h]
   \centering
   \includegraphics[width=8.5cm]{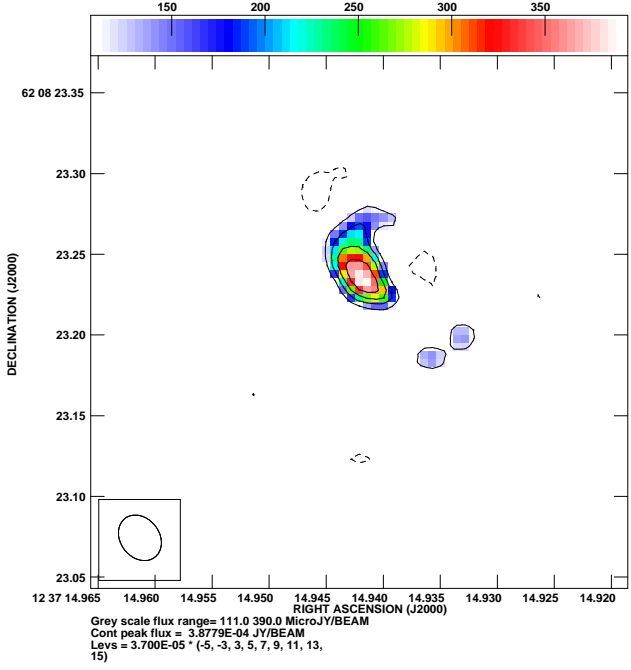}
   \caption{VLA J123714+620823}
   \label{Fig10}
\end{figure}

\subsubsection{VLA J123716+621512}

VLA J123716+621512 has a relatively flat radio spectrum ($\alpha=0.41$)
\citep{R00}.  The radio structure consists of a compact component and
one-sided emission extending $\sim$0.4 arcsec to the south-west
\citep{muxlow05}; the radio structure is hosted by an $I=19.8$ galaxy at
$z=0.561$ (Cowie 2012, priv.comm.).  The VLBI observations detect a
$\sim 0.1$\,mJy compact radio core in this $q_{24} = -0.06$ object. 
Recent VLA observations \citep{M10} indicate radio variability. 

\begin{figure}[h]
   \centering
   \includegraphics[width=8.5cm]{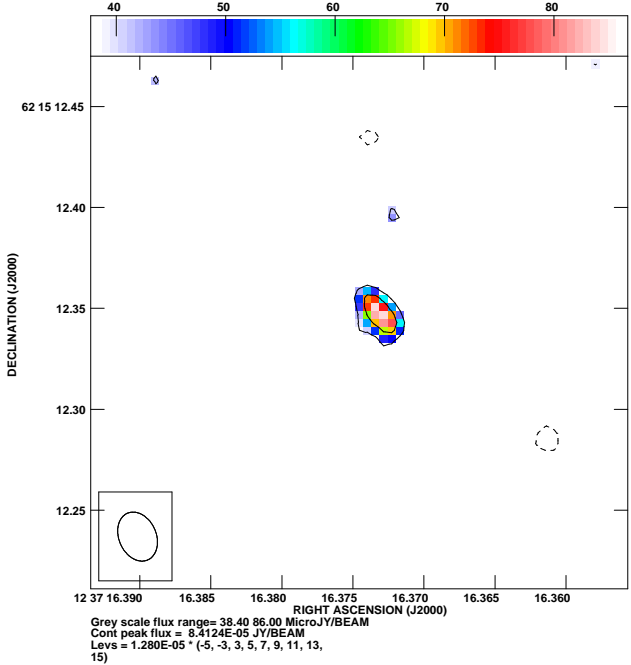}
   \caption{VLA J123716+621512}
   \label{Fig11}
\end{figure}

\subsubsection{VLA J123716+621733}

VLA J123716+621733 is a 0.35mJy radio source with an $I=22.2$ optical
counterpart at a redshift of 1.146 \citep{cowie04}.  The MERLIN-VLA
survey detected this radio source, finding a compact component with a
$\sim 0.6$ arcsec one-sided extension to the south-west
\citep{muxlow05}.  The infrared-bright object was also detected in the
deep {\it Chandra} survey, in both the hard and soft bands.  Our VLBI
detects a compact core containing about half of the VLA flux density. 

\begin{figure}[h]
   \centering
   \includegraphics[width=8.5cm]{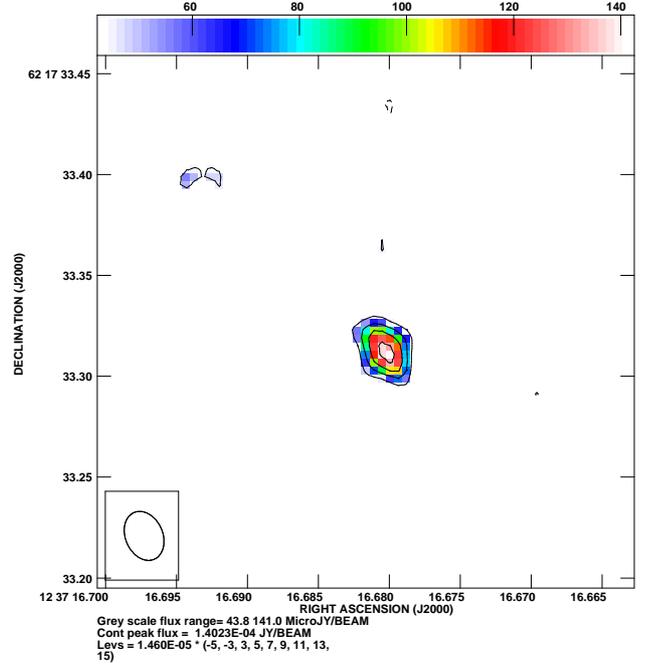}
   \caption{VLA J123716+621733}
   \label{Fig12}
\end{figure}

\subsubsection{VLA J123721+621129}

VLA J123721+621129 has no optical counterpart to $I = 25$ and shows an
inverted radio spectrum ($\alpha = -0.28$) with a dominant compact radio
component and one-sided emission to the north \citep{muxlow05}.  Using
deep imaging data from \citet{barger99}, \citet{richards99} report that
this radio source is associated with a very red object having $I-K >
5.2$.  \citet{barger08} list $z=1.604$.  The compact radio component is
slightly extended at $\sim 25$\,mas resolution, and must be responsible
for most if not all of the radio emission in this moderate radio-excess
($q_{24} = -0.26$) object. 

\begin{figure}[h]
   \centering
   \includegraphics[width=8.5cm]{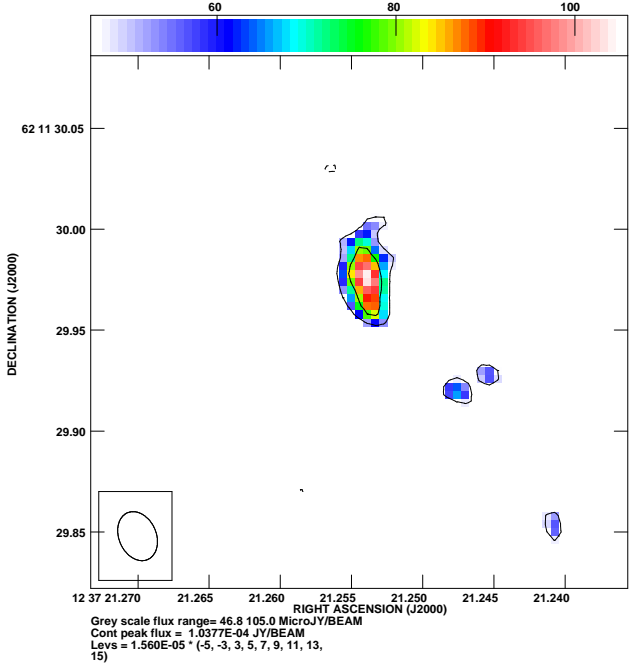}
   \caption{VLA J123721+621129}
   \label{Fig13}
\end{figure}

\section{Discussion}

This research has once more demonstrated the unique strength of high
resolution radio astronomy in assessing the properties, i.e.,
mechanisms-at-work in the distant galaxy population.  Our VLBI study
builds upon the VLA+MERLIN studies of \citet{muxlow05}, confirming,
updating, and refining the results of that study.  Being sensitive to
compact radio emission having brightness temperatures in excess of
$10^5$K, the VLBI observations unambiguously nail down AGN-related
emission.  Combining radio-spectral data with radio morphological and
15$\mu$ infrared information, \citet{muxlow05} classified 18 out of
their sample of 92 faint radio sources as AGN or AGN-candidate (AGNC). 
Their starburst and starburst-candidate class contains 48 objects.  28\%
of the sample, that is 26 objects, remain unclassified.  It is
furthermore very interesting to note that more than half of the total
n=92 sample have {\it Chandra} X-ray counterparts, occurring both in the
starburst and in the AGN class.  With reference to the subfinal column
of Table~1, our VLBI merely detects objects from the VLA+MERLIN AGN and
AGNC classes, and not a single starburst object.  Table~2 summarizes the
VLBI update with respect to the AGN and AGNC classes of
\citet{muxlow05}, and adds the new AGN.  \\

Of the 18 AGN and candidate AGN we confirm six AGN (J123608+621035,
J123644+621133, J123646+621404, J123652+621444, J123714+620823,
J123721+6211295), we nail down the AGN in four AGNC including the submm
source GN16 (J123623+621642, J123700+620909/GN16, J123716+621512,
J123716+621733) and in the $z=4.424$ starburst system with embedded AGN
J123642+621331.  In addition we demonstrate the presence of an AGN in
the $z=0.857$ object J123642+621545.  It is seen that the VLBI misses
out on seven faint ($<150\mu$Jy) AGN/AGNC and on the relatively bright
(5mJy) wide-angle-tail radio galaxy J123725+621128, for the obvious
reasons of sensitivity combined with resolution of extended low surface
brightness emission.  Relevant (total) 1.4GHz radio luminosities of the
11 VLBI detected radio sources with known redshifts, computed adopting a
spectral shape $S_{\nu} \propto {\nu}^{-0.5}$ in the K-correction, range
from $\sim 1 \times 10^{23}$ to $\sim 1 \times 10^{25}$ W/Hz, with
outlier $2 \times 10^{26}$ W/Hz for the starburst with embedded AGN
system J126242+621331 at $z=4.424$.  This is in the range of
radio-luminous Seyfert galaxies and edge-darkened double-lobed
(FR1-type) radio galaxies (NGC\,1068 (3C\,71) and M\,87
(Virgo\,A/3C\,274) have P$_{\rm 1.4GHz} \sim 1 \times 10^{23}$ W/Hz) to
radio-quiet QSOs.  Note that large, double-lobed galaxies (FR2-type) at
$z \sim 1 - 2$ have radio luminosities in excess of $10^{28}$ W/Hz.  The
5\,mJy FR1-type radio galaxy J123725+621128 is identified with an
$I=22.9$ compact galaxy at unknown redshift; assuming $z\sim 1$ its
radio luminosity would be $\sim 10^{25}$ W/Hz.  \\

Considering their infrared/radio ratio $q_{24}$, the VLBI detected
objects have at most a mild radio excess: the AGN generated radio
emission is of similar strength as the starburst generated radio
emission.  Figure~14 illustrates that point.  The diagram compares the
$q_{24}$-values of the 11 VLBI sources with known redshifts with the
$q_{24}$-values of the 24 starburst objects with known redshift from
\citet{muxlow05}.  Both samples contain upper limits, due to {\it
Spitzer} upper limits.  The diagram also specifies average
$q_{24}$-values for the samples (this time including the unknown
redshift objects).  While there is substantial overlap, the AGN stand
out by a moderate $q_{24}$-offset.  The AGN sample displays an average
$\langle q_{24}\rangle=-0.12$, whereas $\langle q_{24}\rangle$ for SBs
is 0.63$\pm$0.43, which is consistent with the previous result of
0.52$\pm$0.37 \citep{beswick08}.  Objects having $q_{24}<0$ are
generally known as radio-excess objects \citep[e.g.][]{donley05}.  It is
seen that four VLBI AGN, having $q_{24}$ = 0.08, 0.55, 0.76 and 1.03,
definitely would not classify as such.  Others are seen to have a mild
excess at most.  Starburst candidate J123646+621629 at $z=0.502$ is seen
to have a radio excess ($q_{24} = -0.7$).  It is puzzling that the VLBI
observations do not detect this object; its MIPS upper limit may be
erroneous or its radio core may be of variable strength (the object is
unique in that sense).  We stress that the integrated (VLA) 1.4\,GHz
flux densities often exceed the VLBI flux densities.  The resolved,
extended radio emission could originate in compact jets or lobes, or
draw from diffuse galactic synchrotron radiation in the AGN host
galaxies.  The substantial 24$\mu$ emission detected in these objects
would argue for an ISM origin in dusty host galaxies.  The SED
decomposition technique involving the NIR, MIR, FIR, submm, and radio
wavebands, recently employed by \citet{moro12}, suggests symbiosis of
dust-obscured star-formation and AGN in several dozens of weak, X-ray
selected objects in the GOODS-N field.  Their radio-excess sample
contains eight of the VLBI AGN which provides confidence in both
techniques.  Surprisingly, a somewhat lower sSFR was established for
these symbiotic objects, in comparison with 24$\mu$ selected
pure-starbursts.  VLBI-undetected (candidate-) AGN from \citet{moro12}
have larger $q_{24}$-values, indicating a smaller AGN contribution. 
Their radio excess flux densities are in the range of $\sim 50$ --
$150{\mu}$Jy.  The required global VLBI sensitivity (over a wide field)
to tackle such radio sources is within reach; we are currently planning
such observations.  A detailed comparison of the results of the SED
decomposition and the radio imaging techniques is beyond the scope of
the present analysis and is postponed to a later paper.  \\

Whereas X-ray selection is a powerful tool to select AGN samples, the
present research underlines its shortcomings.  Two of the VLBI-selected
AGN, J123623+621642 and J123700+620909/GN16 remain without X-ray
detection.  The 2Ms {\it Chandra} observations of \citet{alexander03}
yielded upper limits log\,L$_{\rm 2-10keV} < 42.6$~erg/s for both
objects.  The dusty nature of their host galaxies, referred to above,
could be responsible for X-ray attenuation; this issue again wilt be
dealt with in more detail in a later paper (the X-ray luminosities for
the detected VLBI AGN are in the range $10^{41}$ -- $10^{44}$ erg/s). 
It is in the meantime clear that combination of arcsec scale and
milliarcsec scale radio imaging provides a very powerful and in
principle straightforward tool to select samples of AGN and star-forming
galaxies among the distant galaxy population.  It is no surprise that
such objects are among the main science drivers of the LOFAR and SKA
radio telescopes. \\

To summarize, faint AGN are found in a variety of star-forming galaxies,
with a wide range of dust obscurations, and a relative occurrence of
$\sim 25\%$.  A subsequent paper will address the properties as well as
selection techniques of faint AGN in more detail, and discuss the
general AGN evolution over cosmic time. 

\section{Conclusions}

Deep, high-resolution global VLBI observations at milliarcsec resolution
together with deep arcsec-scale resolution radio imaging provide an
extremely powerful way to separate faint AGN and starburst galaxies, and
assess their symbiotic occurrence among the distant galaxy population. 
Using these techniques on the Hubble Deep Field North and its flanking
fields, it is found that about 25\% of star-forming galaxies of various
types harbour faint AGN.  Some classify as starbursts, such as the
$z=4.42$ system J123642+621331 and the $z=1.68$ submm galaxy GN16/
J123700+620909, and some can be X-ray obscured. 

\begin{acknowledgements}

This paper is dedicated to the memory of Seungyoup Chi, who, coming from
South Korea, worked as promising PhD student in Groningen and Dwingeloo,
but passed away in the final year of his PhD studies.  Authors PDB and
MAG are grateful for communications with Len Cowie and Ranga-Ram Chary,
for discussions with Dave Alexander and Agnese del Moro, and for help
with the graphics by Seyit Hocuk. Comments by an expert referee are also
gratefully acknowledged.

\end{acknowledgements}

\bibliographystyle{aa}
\bibliography{gg053rev3}

\begin{figure*}[!h]
    \resizebox{\hsize}{!}{\includegraphics[clip=true]{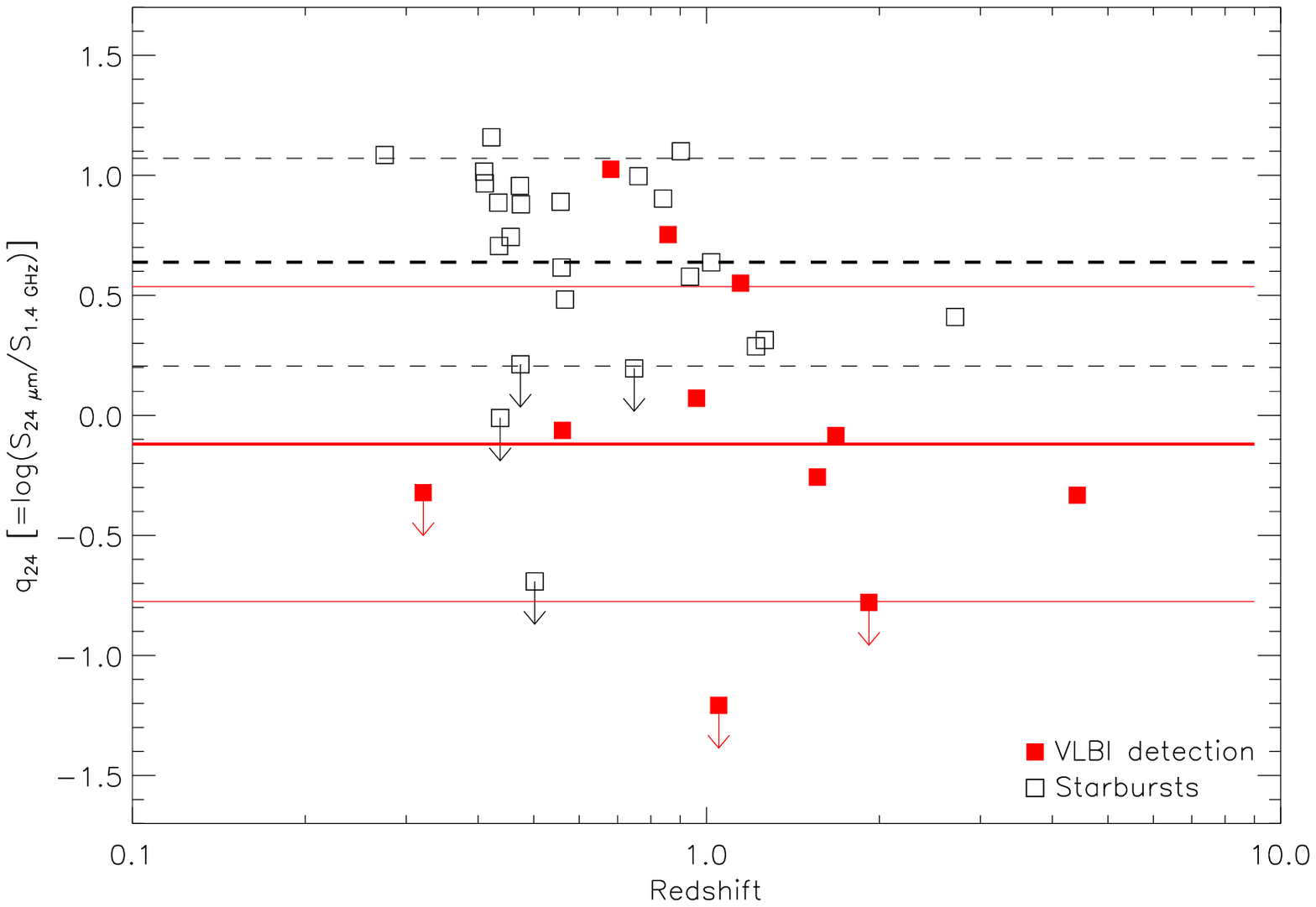}}
    \caption{\footnotesize
      Values of $q_{24}$ for SBs (black empty squares) and 11 global VLBI detections
      (red filled squares), as function of redshift. The thick dashed (black) and solid
      (red) lines represent the mean values for each sample, respectively, and the
      1$\sigma$ deviation is shown by thinner black dashed and red lines.}
    \label{fig14}
\end{figure*}

\begin{landscape}
\addtolength{\headsep}{5.4cm}
\begin{table*}
\caption{Summary: compact radio sources in HDF-N and HFF.}
\label{table1} \centering
\begin{tabular}{lccccccccclr}
\hline\hline
Name & R.A. & Dec. & $z$ & $S_{VLA}$$~^{a}$ & $S_{WSRT}$$~^{a}$ & $S_{Global\,VLBI}$ & $S_{24\mu m}$$~^{b}$ & $F_{X}$$~^{c}$ & Radio & Classification & $q_{\rm 24}$ \\
~ &  [hh mm ss] &  [+dd mm ss] & ~ & [$\mu$Jy] & [$\mu$Jy] & [$\mu$Jy] & [$\mu$Jy] & [$10^{-18}$Wm$^{-2}$] & structure & ~ & ~ \\
\hline
  \multicolumn{10}{c}{Inner HDF-N and adjacent HFF}\\ 
\hline
  J123642+621331 (core) & 12 36 42.0908 & 62 13 31.425 & \multirow{2}{*}{4.424$~^{e}$} & \multirow{2}{*}{467} & \multirow{2}{*}{489/489} &
  227$\pm$26 & \multirow{2}{*}{209$\pm$5.59} & \multirow{2}{*}{0.28}& \multirow{2}{*}{C1E} & \multirow{2}{*}{SB+AGN} & \multirow{2}{*}{$-0.35$} \\
  J123642+621331 (jet) & ~ & ~ & ~ & ~ & ~ & 121$\pm$21 & ~ & ~ \\
  J123644+621133 & 12 36 44.3870 & 62 11 33.145 & 1.050$~^{f}$ & 1290 & 1190/1606 & 309$\pm$27 & $\leq 80$ & 0.24 & FR1 & AGN (FR1) & $\leq -1.21$ \\
  J123646+621404 & 12 36 46.3321 & 62 14 04.693 & 0.961$~^{g}$ & 179 & 187/187 & 247$\pm$14 & 213$\pm$5.22 & 24.6 & C1E & AGN & $0.08$ \\
  J123652+621444 & 12 36 52.8839 & 62 14 44.076 & 0.321$~^{g}$ & 168 & 237/276 & 83$\pm$13 & $\leq 80$ & 0.46 & C1E & AGN & $\leq -0.32$ \\
\hline
  \multicolumn{10}{c}{Outer HFF} \\
\hline
  J123608+621035 & 12 36 08.1195 & 62 10 35.898 & 0.681$~^{g}$ & 217 & 190/246 & 140$\pm$30 & 2300$\pm$13.50 & 1.89 & CE & AGN & $1.03$ \\
  J123623+621642 & 12 36 23.5436 & 62 16 42.754 & 1.918$~^{h}$ & 481 & 476/476 & 327$\pm$50 & $\leq 80$ & $-$ & C1E & AGN candidate & $\leq -0.78$ \\
  J123642+621545 & 12 36 42.2123 & 62 15 45.521 & 0.857$~^{i}$ & 150 & 88/89 & 343$\pm$101 & 866$\pm$10.30 & 2.46 & CE & U (SB+AGN) & $0.76$ \\
  J123700+620909 & 12 37 00.2480 & 62 09 09.778 & 1.68$~^{j}$ & 324 & 236/236 & 147$\pm$34 & 267$\pm$7.62 & $-$ & C1E & AGN candidate (GN16) & $-0.08$ \\
  J123714+620823 & 12 37 14.9414 & 62 08 23.208 & $(0.847)~^{k} $ & 1350 & 1853/1853 & 645$\pm$80 & 215$\pm$6.67 & 0.6 & C & AGN & $-0.80$ \\
  J123716+621512 & 12 37 16.3740 & 62 15 12.343 & 0.561$~^{l}$ & 187 & 166/166 & 110$\pm$25 & 162$\pm$9.40 & 0.23 & C1E & AGN candi. (AGN) & $-0.06$ \\
  J123716+621733 & 12 37 16.6811 & 62 17 33.327 & 1.146$~^{g}$ & 346 & 362/362 & 177$\pm$25 & 1240$\pm$15.50 & 21.69 & C1E & AGN candi. (AGN) & $0.55$ \\
  J123721+621129 & 12 37 21.2539 & 62 11 29.954 & 1.56$~^{m}$ & 383 & 381/381 & 254$\pm$51 & (210$\pm$6.29)$~^{d}$ & 0.51 & C1E & AGN & $-0.26$ \\
\hline
\\
  \multicolumn{10}{l}{{\footnotesize $^{a}$ \citet{muxlow05}; the WSRT entries represent peak/total values}} \\
  \multicolumn{10}{l}{{\footnotesize $^{b}$ {\it Spitzer}: merged IRAC and MIPS GOODS-N catalog (R.~Chary -- 2011 private communication)}} \\
  \multicolumn{10}{l}{{\footnotesize $^{c}$ \citet{alexander03, donley05}}} \\
  \multicolumn{10}{l}{{\footnotesize $^{d}$ $\sim 1.5 ''$ offset from MIPS position, $\sim 1 ''$ offset from IRAC}} \\
  \multicolumn{10}{l}{{\footnotesize $^{e}$ \citet{waddington99}}} \\
  \multicolumn{10}{l}{{\footnotesize $^{f}$ \citet{cohen00}}} \\
  \multicolumn{10}{l}{{\footnotesize $^{g}$ \citet{cowie04}}} \\
  \multicolumn{10}{l}{{\footnotesize $^{h}$ \citet{chapman04}}} \\
  \multicolumn{10}{l}{{\footnotesize $^{i}$ \citet{hornschemeier01}}} \\
  \multicolumn{10}{l}{{\footnotesize $^{j}$ \citet{pope06}}} \\
  \multicolumn{10}{l}{{\footnotesize $^{k}$ L.~Cowie -- 2010 private communication: tentative value}} \\
  \multicolumn{10}{l}{{\footnotesize $^{l}$ L.~Cowie -- 2012 private communication; not to be confused with J123716+621511 at $z=0.231$}} \\
  \multicolumn{10}{l}{{\footnotesize $^{m}$ \citet{barger02}}} \\

\end{tabular}
\end{table*}
\end{landscape}

\begin{table*}
\caption{VLBI detected and undetected AGN and AGN candidates from the VLA+MERLIN 
 1.4\,GHz survey \citep{muxlow05}. Note that the VLBI in addition detected the
 starburst+AGN system J123642+621331 at $z=4.42$ and the unclear identification
 J123642+621545 at $z=0.857$. The VLBI sensitivity ranges from $\sim50\mu$Jy in
 the field center to $\sim150\mu$Jy at r$_{\rm phase center} >$~6~arcmin (see text).
}
\label{table2} 
\centering
\begin{tabular}{rrrrrr}
\hline\hline
Cat.number$^{a}$ & Name$^{b}$ & Alt.name & 1.4\,GHz VLA flux density & VLA class$^{c}$ & VLBI detection \\
\hline
4  & J123608+621035 & ~    &  217$\mu$Jy  & AGN  & yes  \\
15 & J123618+621635 & ~    &   47$\mu$Jy  & AGNC & no   \\
18 & J123620+620844 & ~    &  123$\mu$Jy  & AGN  & no   \\
21 & J123622+621544 & ~    &   84$\mu$Jy  & AGNC & no   \\
24 & J123623+621642 & ~    &  481$\mu$Jy  & AGNC & yes  \\
39 & J123640+621009 & ~    &   87$\mu$Jy  & AGNC & no   \\
43 & J123644+621133 & ~    & 1290$\mu$Jy  & AGN  & yes  \\
47 & J123646+621404 & ~    &  179$\mu$Jy  & AGN  & yes  \\
56 & J123652+621444 & ~    &  168$\mu$Jy  & AGN  & yes  \\
64 & J123700+620909 & GN16 &  324$\mu$Jy  & AGNC & yes  \\
72 & J123709+620837 & ~    &   72$\mu$Jy  & AGN  & no   \\
73 & J123709+620841 & ~    &   68$\mu$Jy  & AGNC & no   \\
74 & J123711+621330 & ~    &  132$\mu$Jy  & AGNC & no   \\
78 & J123714+620823 & ~    & 1350$\mu$Jy  & AGN  & yes  \\
79 & J123716+621512 & ~    &  187$\mu$Jy  & AGNC & yes  \\
81 & J123716+621733 & ~    &  346$\mu$Jy  & AGNC & yes  \\
85 & J123721+621129 & ~    &  383$\mu$Jy  & AGN  & yes  \\
89 & J123725+621128 & ~    &     5mJy     & AGN  & no   \\
\hline
\\
  \multicolumn{5}{l}{{\footnotesize $^{a}$ number in VLA/MERLIN catalog \citep{muxlow05}, at increasing RA}} \\
  \multicolumn{5}{l}{{\footnotesize $^{b}$ J2000 name of radio source \citep{muxlow05} }} \\
  \multicolumn{5}{l}{{\footnotesize $^{c}$ the \citet{muxlow05} classification combines radio and
                           infrared ({\it ISO}) information}} \\

\end{tabular}
\end{table*}

\end{document}